\documentclass[superscriptaddress,twocolumn,10pt]{revtex4-1}

\usepackage{amsmath}    
\usepackage{graphicx}   
\usepackage{verbatim}   
\usepackage{color}      
\usepackage{subfigure}  
\usepackage{hyperref}   
\usepackage{amssymb}
\usepackage{amsthm}
\usepackage{amsfonts}
\begin{document}


\title{Modulus--Pressure Equation for Confined Fluids} 

\author{Gennady Y. Gor} \email[Corresponding author, e-mail: ]{gennady.y.gor@gmail.com} 
\affiliation{NRC Research Associate, Resident at Center for Materials Physics and Technology, Naval Research Laboratory, Washington, DC 20375, USA}
\author{Daniel W. Siderius}
\author{Vincent K. Shen}
\affiliation{Chemical Sciences Division, National Institute of Standards and Technology, Gaithersburg, MD 20899, USA}
\author{Noam Bernstein}
\affiliation{Center for Materials Physics and Technology, Naval Research Laboratory, Washington, DC 20375, USA}

\date{\today}

\begin{abstract}
\noindent Ultrasonic experiments allow one to measure the elastic modulus of bulk solid or fluid samples. Recently such experiments have been carried out on fluid-saturated nanoporous glass to probe the modulus of a confined fluid. In our previous work [J. Chem. Phys., (2015) {\bf143}, 194506], using Monte Carlo simulations we showed that the elastic modulus $K$ of a fluid confined in a mesopore is a function of the pore size. Here we focus on modulus-pressure dependence $K(P)$, which is linear for bulk materials, a relation known as the Tait-Murnaghan equation.
Using transition-matrix Monte Carlo simulations we calculated the elastic modulus of bulk argon as a function of pressure and argon confined in silica mesopores as a function of Laplace pressure. Our calculations show that while the elastic modulus is strongly affected by confinement and temperature, the slope of the modulus versus pressure is not. Moreover, the calculated slope is in a good agreement with the reference data for bulk argon and experimental data for confined argon derived from ultrasonic experiments. We propose to use the value of the slope of $K(P)$ to estimate the elastic moduli of an unknown porous medium. 


\end{abstract}

\maketitle

\section{Introduction}

When fluids are confined to nanometer-scale pores, their thermodynamic and dynamic properties can significantly differ from those of bulk fluids~\cite{Huber2015}. Simple examples of these changes are shifts of the temperatures and pressures at which phase transitions occur~\cite{Morishige1999, Wallacher2001, Morishige2012}, and the development of extremely high pressures~\cite{Gunther2008, Long2012, Gor2013}, etc. 
Experimental studies of properties of confined fluids often presents a challenge, and in recent years much effort has been spent employing various techniques, primarily X-ray~\cite{Morineau2003, Gunther2008} and neutron scattering~\cite{Melnichenko2015}.  Interestingly, very short wavelength tools are not the only ones able to shed light on the nano-confined fluid properties, but also long wavelength tools, such as ultrasonic acoustic waves. 

In 1982 Murphy carried out the first ultrasonic study on a fluid-saturated Vycor glass, which was focused mainly on sound attenuation~\cite{Murphy1982}. This work introduced the use of ultrasound for studies of properties of matter in nanometer confinement of Vycor pores, which has been employed in numerous works since then. Beamish and co-workers used it to study the properties of liquid helium~\cite{Beamish1983, Beamish1984, Mulders1989} and also proposed ultrasonic experiments for probing the surface area of nanoporous materials~\cite{Warner1988}. Later ultrasonic experiments became widely used for studying phase transitions in confinement, both liquid-vapor~\cite{Page1993, Page1995, Schappert2013JoP, Schappert2014} and solid-liquid~\cite{Molz1993, Molz1995, Beaudoin1996, Charnaya2001, Borisov2006, Schappert2008, Charnaya2008, Borisov2009, Schappert2011, Schappert2013PRL}. Among these works, it is worth pointing out the paper of Page et al.~\cite{Page1995}, which first showed that when the pores are completely filled with a liquid-like capillary condensate, the modulus of the fluid  (n-hexane) was not constant, but a monotonic function of the gas pressure $p$ (lower pressure corresponds to lower modulus). Moreover, they explained this regularity by the negative Laplace pressures due to menisci at the liquid-vapor interfaces below saturation. More recently Schappert and Pelster carried out a similar experiment, but using argon at cryogenic temperatures~\cite{Schappert2014}, and their results were consistent with the results of Page et al. 

These studies motivated us to perform calculations of the elastic modulus of confined argon using classical density functional theory (cDFT)~\cite{Gor2014} and Monte Carlo simulations~\cite{Gor2015compr}. Our calculations based on cDFT reproduced the experimentally observed  linear relation between the modulus and the solvation pressure (also referred to as the adsorption stress), and the slope in this relation was close to the slope reported in ultrasonic experiments~\cite{Schappert2014}. However, we did not compare this result to the bulk data. Here we present simulation results for the elastic modulus of liquid argon at several different temperatures, both in bulk and under confinement in nanopores of different sizes and solid-fluid interaction strengths. We show that for all the simulated cases, the elastic modulus is a linear function of fluid pressure (in bulk) or Laplace pressure (in confinement) {\em with the same slope}. Moreover, we show that this slope agrees well with the slope calculated from experimental reference data for bulk liquid argon. Finally, we show  that this slope is close to the slope calculated from the ultrasonic experiments of Schappert and Pelster. These comparisons demonstrate that the slope of the $K(P)$ dependence is roughly constant irrespective of the confinement. To our knowledge, this common slope has not been reported previously for confined fluids. In addition to contributing to fundamental understanding of thermodynamics of confined fluids, this common slope has useful implications for ultrasonic studies of porous materials: we discuss how our result can be applied for estimating the elastic parameters of nanoporous solids from ultrasonic experiments.

\section{Methods}
\label{sec:methods}

\subsection{Elastic Properties from Ultrasonic Experiments}
\label{sec:ultrasonic}

Experimental measurements of the propagation time of ultrasonic waves give information on its velocity, which provides the information for elastic constants of materials: shear modulus $G$ from the shear wave velocity $v_s$, and longitudinal modulus $M$ from the longitudinal wave velocity $v_l$:
\begin{equation}
\label{waves}
G = v_s^2 \rho ~~~~~~~ M = v_l^2 \rho,
\end{equation}
where $\rho$ is the mass density of the sample. The bulk modulus of the sample can be calculated from the two measurements as~\cite{Goodman1989} 
\begin{equation}
\label{bulk-modulus}
K = M - \frac{4}{3} G.
\end{equation}
For fluids, the shear modulus $G_f$ is zero and therefore the longitudinal modulus is equal to the bulk modulus
\begin{equation}
\label{bulk-fluid}
K_f = M_f.
\end{equation}

\begin{figure}[ht]
\centering
\includegraphics[width=0.95\linewidth]{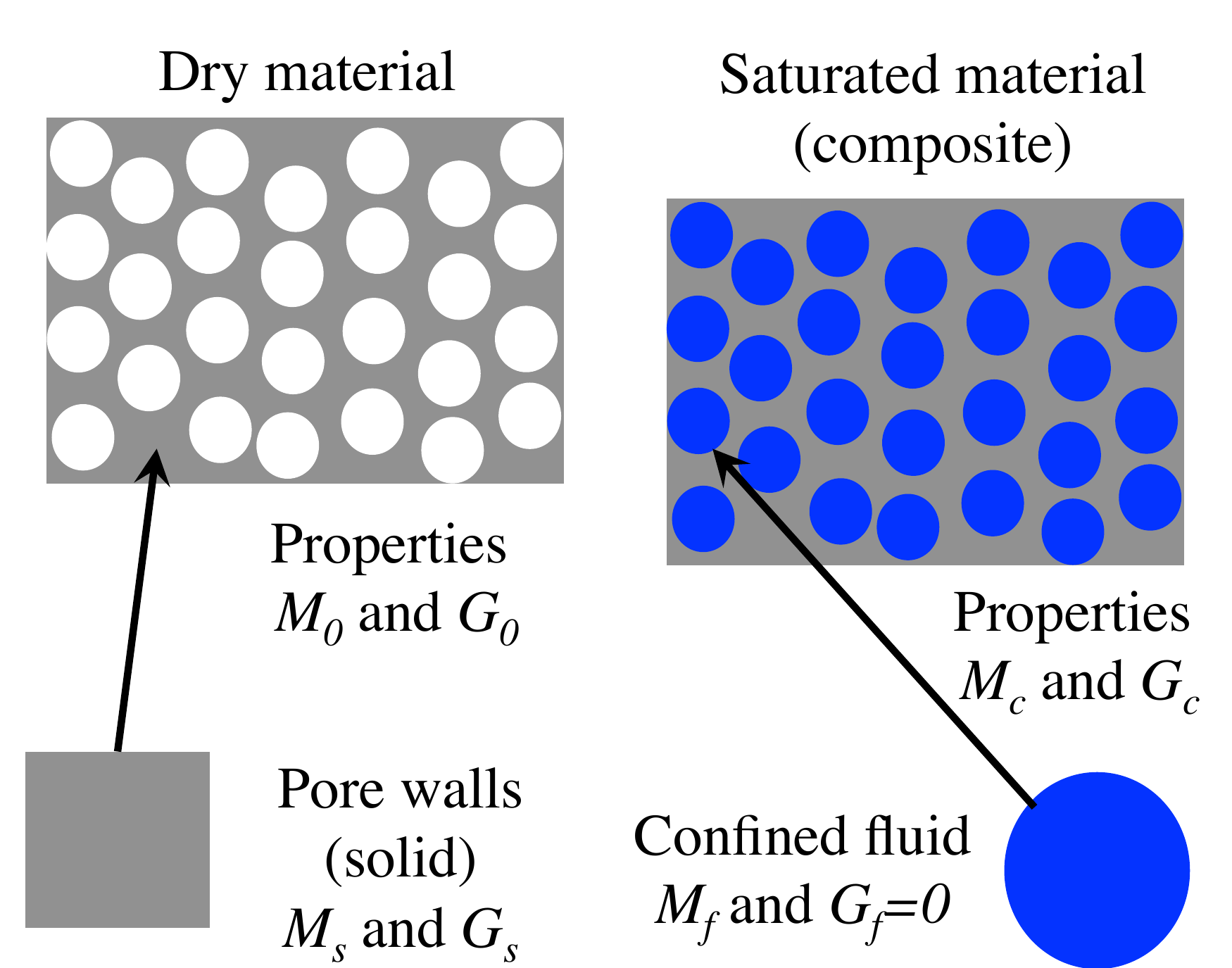} \\
\caption{Schematic of a fluid-saturated porous material as a composite, with denotations for various elastic moduli.}
\label{fig:Porous}
\end{figure} 
 
While the calculation of a fluid bulk modulus from ultrasonic experiments is straightforward, the calculation of the modulus of a confined fluid from the experimental ultrasonic data is non-trivial. For a given porous sample, one can measure the logitudinal and shear moduli when it is dry ($M_0$ and $G_0$) and when it is saturated, i.e., a solid-fluid composite ($M_c$ and $G_c$). The differences between the corresponding moduli are determined by the properties of the fluid. Experiments for argon adsorption in Vycor glass at $T = 80$~K and $T = 86$~K show that the shear modulus of the sample does not change, $G_c \simeq G_0$~\cite{Schappert2014}, which is consistent with the fact the adsorbed argon is in a liquid-like state and the shear modulus of the liquid is zero. 

Experiments showed that in contrast with the constant shear modulus, the longitudinal modulus of the saturated sample $M_c$ is several percent higher than the longitudinal modulus of the dry material $M_0$~\cite{Schappert2014}. Moreover, for the argon-filled pores the change of the longitudinal modulus of the sample $\Delta M = M_c - M_0$ is a function of equilibrium gas pressure $p$. Assuming that the properties of the solid do not change appreciably, this variation can be attributed to the change of the modulus of the confined fluid. Schappert and Pelster proposed an effective medium model to determine the longitudinal modulus of the fluid $M_f$ from the change of the longitudinal modulus of the sample $\Delta M$~\cite{Schappert2013JoP}
\begin{equation}
\label{EM}
M_f = C \Delta M,
\end{equation}
where the constant $C$ is a function of parameters of the dry material (porous matrix) $M_0$, $G_0$ and pore walls (constituent solid) $M_s$, $G_s$ (see Figure~\ref{fig:Porous}). Interestingly, Page et al.~\cite{Page1995} used a similar proportionality relation, where the constant $C$ was a function of different material parameters,  based on works of Gassmann and Biot~\cite{Gassmann1951, Biot1956i}. The discussion of the constant $C$ is beyond the scope of the current work, but it is important to note that both approaches suggest that there should be a \textit{direct proportionality} between the measured $\Delta M$ and $M_f$. Finally, taking into account $G_f = 0$ we can interpret the resulting longitudinal modulus as the bulk modulus of the fluid, i.e.,\ $K_f = M_f$.

\subsection{Thermodynamic Definitions of Elastic Properties}
\label{sec:thermo}

In thermodynamics the most common property that quantifies the elastic properties of a fluid is the isothermal compressibility $\beta_T$, defined as
\begin{equation}
\label{beta-def}
\beta_T \equiv - \frac{1}{V} \left( \frac{\partial V}{\partial P}\right)_{T},
\end{equation}
where $V$ is the volume of the fluid, $P$ is the fluid pressure, and $T$ is the absolute temperature. The isothermal compressibility of a bulk or confined fluid can be straightforwardly calculated from molecular simulations (See Section~\ref{sec:simulations}). For further consideration and for comparison with ultrasonic experiments it is convenient to discuss the reciprocal value, the isothermal elastic modulus 
\begin{equation}
\label{Kt-def}
K_T = \frac{1}{\beta_T} \equiv - V \left( \frac{\partial P}{\partial V}\right)_{T}.
\end{equation}
It is important to note that the modulus $K_T$ is not the experimental bulk modulus determined from ultrasonic experiments. The experimental conditions are adiabatic rather than isothermal. Therefore the experimentally-measured bulk modulus is actually the adiabatic modulus, i.e., determined by a derivative at constant entropy $S$:
\begin{equation}
\label{Ks-def}
K \equiv - V \left( \frac{\partial P}{\partial V}\right)_{S}.
\end{equation}
However, $K$ and $K_T$ are simply related through the heat capacity ratio $\gamma \equiv c_P/c_V$~\cite{Landau5} via
\begin{equation}
K_T = K/\gamma.
\end{equation}
Ultrasonic and calorimetric measurements for bulk fluids are well-established, and thus the values of compressibilities (or moduli) for many bulk fluids are readily available for a wide range of temperatures and pressures. For argon, which is considered in this work, the reference data is given in~\cite{Stewart1989, Tegeler1999, Linstrom_NIST_WebBook}.

Strictly speaking, in order to relate $K_T$ and $K$ for a confined fluid, one needs to know the heat capacity ratio $\gamma$ for the \textit{confined fluid}. However, it is unlikely that confinement has a noticeable effect on the heat capacity of argon~\cite{Huber1999, Wallacher2001, Knorr2003}, and moreover on the heat capacity \textit{ratio} $\gamma$. Therefore, we use the bulk value of $\gamma$ at the corresponding temperature when converting between $K_T$ and $K$ for the confined fluid. Note also that confinement introduces anisotropy to the fluid, and unlike the bulk fluid it is not always sufficient to describe the pressure in terms of a scalar variable. Calculations of a pressure tensor for simple fluids in confinement show differences between the normal and tangential components \cite{Brodskaya2010i,Long2013Colloids}. Our previous calculations showed that the compressibility of a confined fluid should be related to the normal component of the pressure tensor.

\subsection{Calculating Elastic Properties from Molecular Simulations}
\label{sec:simulations}

Classical statistical mechanics can be used to calculate the compressibility of the fluid $\beta_T$ from the fluctuations of number of particles in the pore, $N$, in the grand-canonical ensemble~\cite{Landau5, Allen1989} through the relation
\begin{equation}
\label{beta-fluct}
K_T^{-1} = \beta_T = \frac{V \langle \delta N^2 \rangle}{k_{\rm{B}} T \langle N\rangle^2 },
\end{equation}
where $\langle \delta N^2 \rangle = \langle N^2 \rangle - \langle N \rangle^2$ and $k_{\rm{B}}$ is Boltzmann's constant. Eq.~\ref{beta-fluct} is valid when the fluctuations are normally distributed~\cite{Landau5}. In the grand-canonical ensemble the fluid atoms or molecules in the pore are assumed to be in equilibrium with a reservoir, and the pressure $p$ of the reservoir is the pressure of a bulk fluid at the same chemical potential and $T$ as the fluid in the pore. Typically, the reservoir contains a gas phase whose pressure, $p$, differs from the pressure in the adsorbed phase, $P$.

Here we use results from molecular simulations of argon in spherical silica pores from the grand canonical transition-matrix Monte Carlo (GC-TMMC)~\cite{Errington_Direct_2003, Shen_Metastability_2004,Siderius2013} method. Since this method was discussed and used in previous work~\cite{Gor2015compr}, the reader is directed to Section~II.B~of Ref.~\onlinecite{Gor2015compr} for a full discussion of both the simulation technique and the fluid model. In short, we modeled argon at various temperatures below its critical point both as a bulk fluid and confined in model spherical mesopores with simulation strategies identical to those in Ref.~\onlinecite{Gor2015compr}. The parameters used to model the argon fluid and silica material were identical to those in Ref.~\onlinecite{Gor2015compr}, except for simulations which varied the solid-fluid interaction strength. In the base case of argon-silica, the solid-fluid interaction parameter is $\epsilon_{\rm{sf}}/k_B = 171.24~K$; to investigate the effect of varying this parameter we ran simulations at $\epsilon_{\rm{sf}}/k_B$ $\pm 20\%$, $\pm 50\%$ relative to the base case.

For the discussion we choose pore sizes 3.0~nm, 4.0~nm and 5.0~nm. By the pore sizes we refer here to the external diameter of a spherical pore $d_{\mathrm{ext}}$, the distance of a line drawn through the centers of hypothetical silica solid atoms at opposite pore wall surfaces. The internal diameter is given by $d_{\mathrm{int}} \simeq d_{\mathrm{ext}} - 1.7168 \sigma_{\rm{sf}} + \sigma_{\rm{ff}}$, which well approximates the accessible volume of the pore~\cite{Gor2012}. Lastly, we present results at temperatures of 77.7~K, 87.3~K, and 95.7~K; the results for confined argon at 87.3~K are taken directly from Ref.~\onlinecite{Gor2015compr}, the results for bulk at 87.3~K and all the results at the other temperatures are from new simulations.

\section{Results}
\label{sec:results}

\begin{figure}[ht]
\centering
\includegraphics[width=0.95\linewidth]{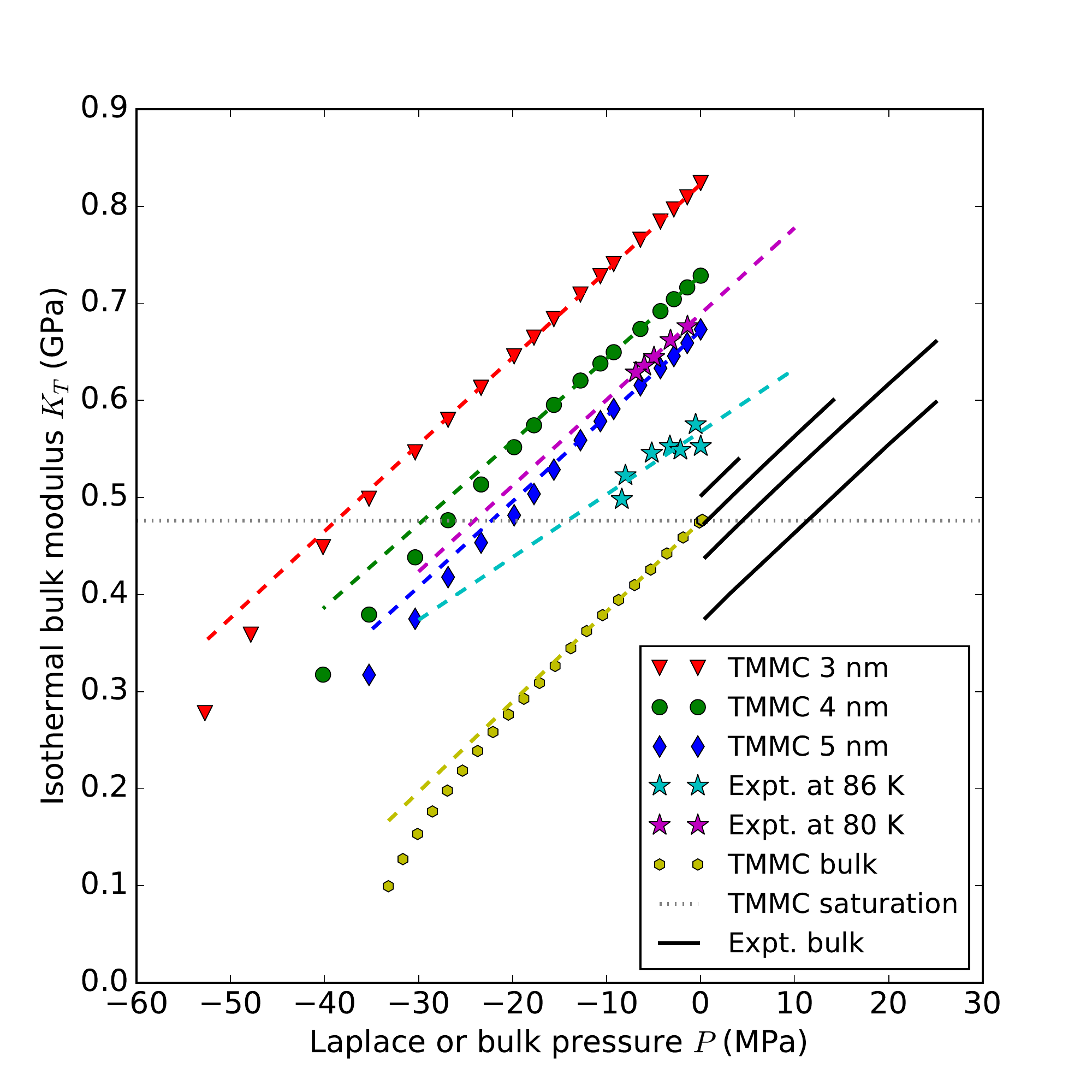} \\
\caption{Liquid argon isothermal elastic modulus $K_T$ as a function of pressure $p$ or Laplace pressure $P_L$ for bulk and confined systems, respectively. Black solid lines show the experimental reference data for bulk argon at temperatures of 85~K, 87.3~K, 90~K and 95~K (top to bottom), from Ref.~\onlinecite{Tegeler1999}. The stars are the moduli calculated based on the ultrasonic experiments for confined argon from Refs.~\onlinecite{Schappert2014} and \onlinecite{Schappert2014Langmuir}, other symbols are the results of GC-TMMC calculations, and dashed lines show the linear fit for each series. The horizontal dotted line indicates $K_T$ of bulk argon at saturation point at 87.3~K calculated by GC-TMMC.}  
\label{fig:Argon}
\end{figure}

GC-TMMC simulations provide the compressibility (or elastic modulus) of the fluid as a function of the relative gas pressure $p/p_0$, where $p_0$ is the saturation pressure for the fluid at the specified temperature. At $p_c < p < p_0$, where $p_c$ is the capillary condensation pressure, a mesopore is filled with a liquid-like capillary condensate. Such system is under the action of capillary (Laplace) pressure, which causes stretching of the fluid. Note that the tension of the liquid in the mesopores is clearly seen in the X-ray diffraction patterns~\cite{Morishige2006} and through the adsorption-induced deformation of the mesopores~\cite{Gunther2008, Gor2013}. The Laplace pressure can be written as a function of $p/p_0$ using the Kelvin-Laplace equation
\begin{equation}
\label{Laplace}
P_L = \frac{R_g T}{V_l} \ln\left(\frac{p}{p_0}\right),
\end{equation}
where $R_g$ is the gas constant and $V_l$ is the liquid molar volume. Since $V_l$ varies only slightly with $p/p_0$, for simplicity we use $V_l$ corresponding to the bulk liquid at saturation conditions in the results that follow. For $p/p_0 < 1$, $P_L$ is necessarily negative and is indicative of the stretched nature of the fluid. Therefore, plots of the calculated modulus $K_T$ for the confined fluid versus Laplace pressure show the variation of the fluid's elastic properties as a function of the actual pressure in the adsorbed phase. Figure~\ref{fig:Argon} gives these dependences for three modeled confined systems: 3~nm, 4~nm and 5~nm pores at 87.3~K (GC-TMMC results from Ref.~\onlinecite{Gor2015compr}).

\begin{figure}[ht]
\centering
\includegraphics[width=0.95\linewidth]{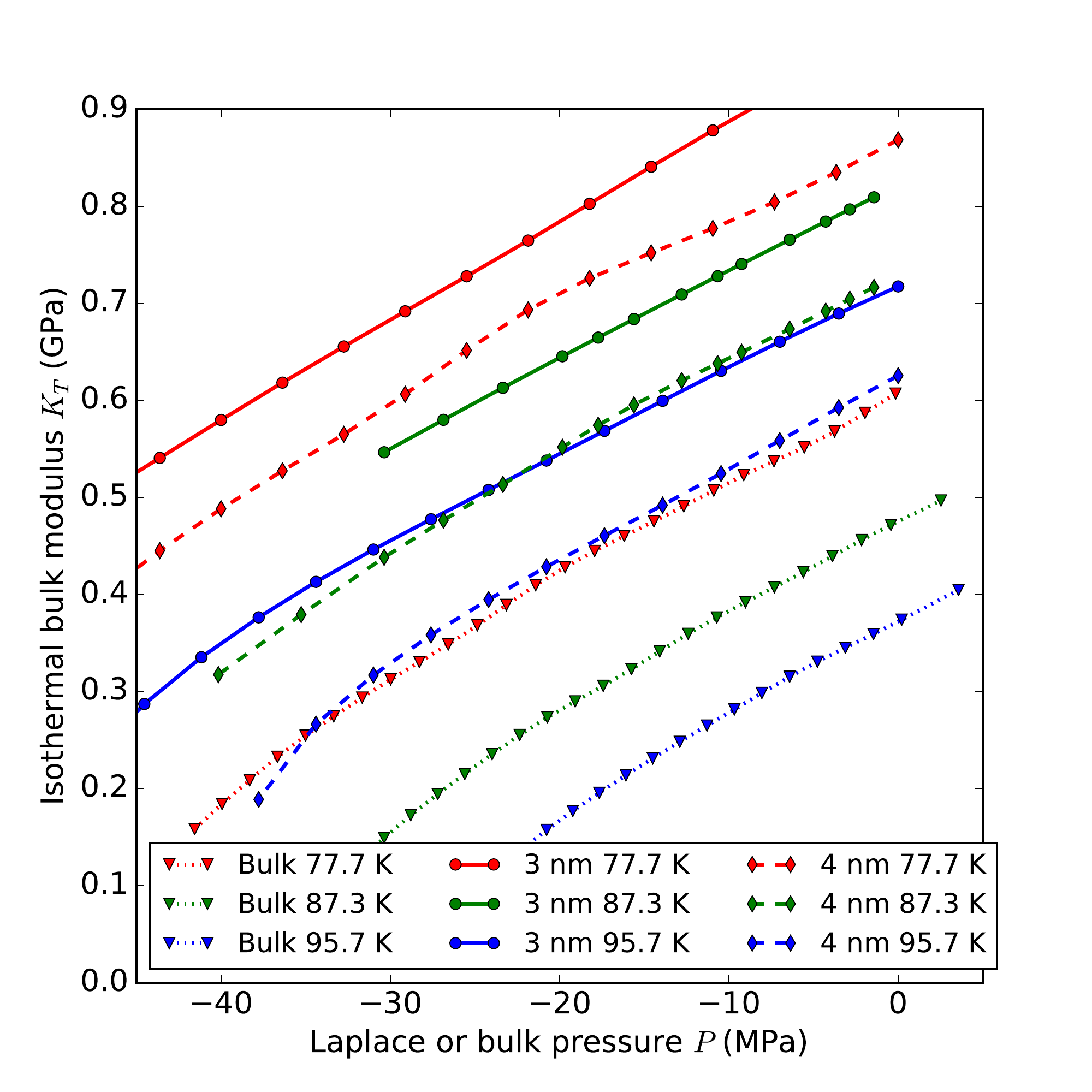} \\
\caption{Simulation results for the isothermal elastic modulus of liquid argon $K_T$ as a function of pressure $p$ for bulk system the pressure, and as a function of Laplace pressure $P_L$ for confined systems with noted pore sizes, at temperatures of 77.7~K, 87.3~K, and 95.7~K.}
\label{fig:Temperature}
\end{figure}

\begin{figure}[ht]
\centering
\includegraphics[width=0.95\linewidth]{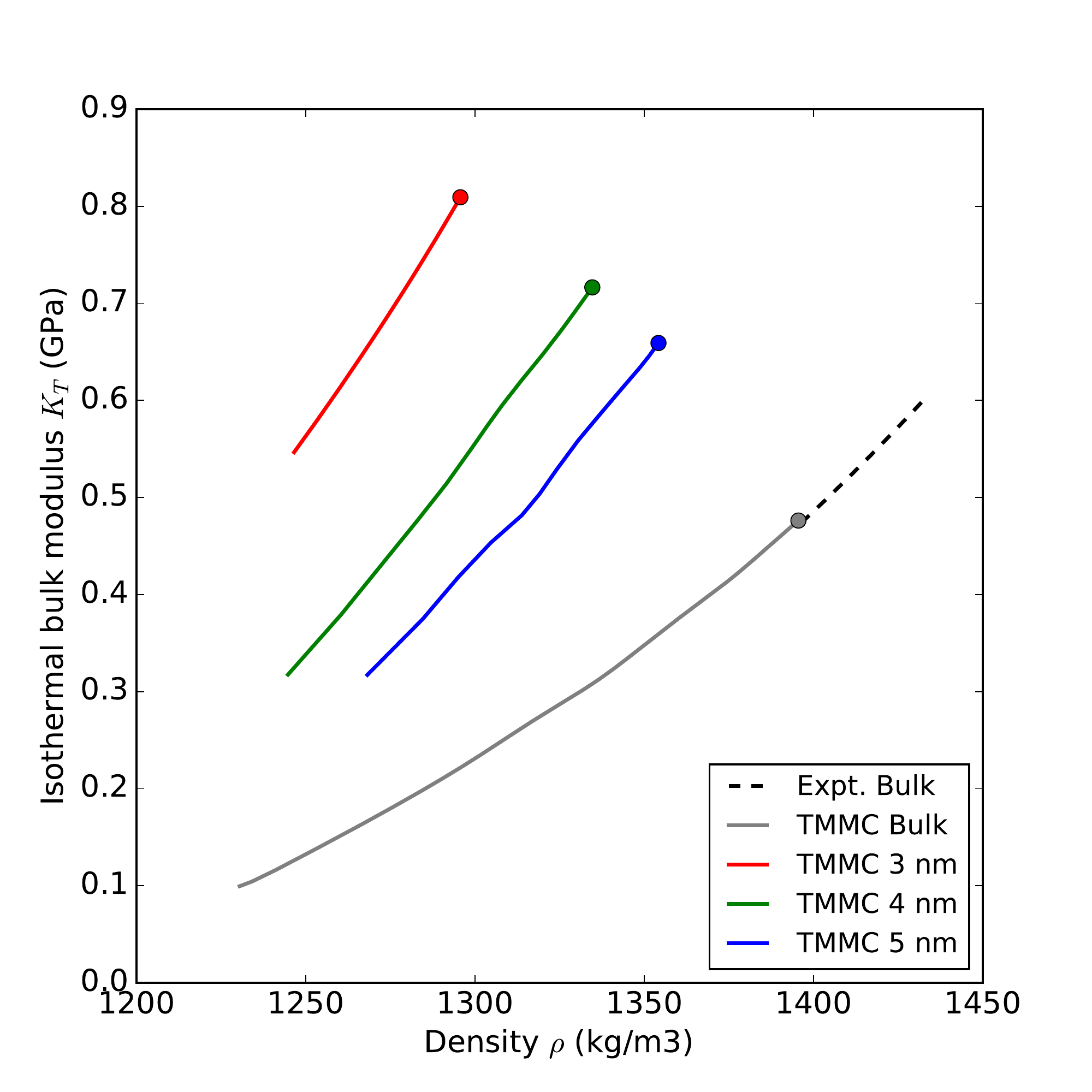} \\
\caption{Isothermal elastic modulus of liquid argon $K_T$ at 87.3~K as a function of density in bulk (experimental data and simulations) and confined fluid (simulations). The markers correspond to the saturation points.}
\label{fig:Density}
\end{figure} 

\begin{figure}[ht]
\centering
\includegraphics[width=0.95\linewidth]{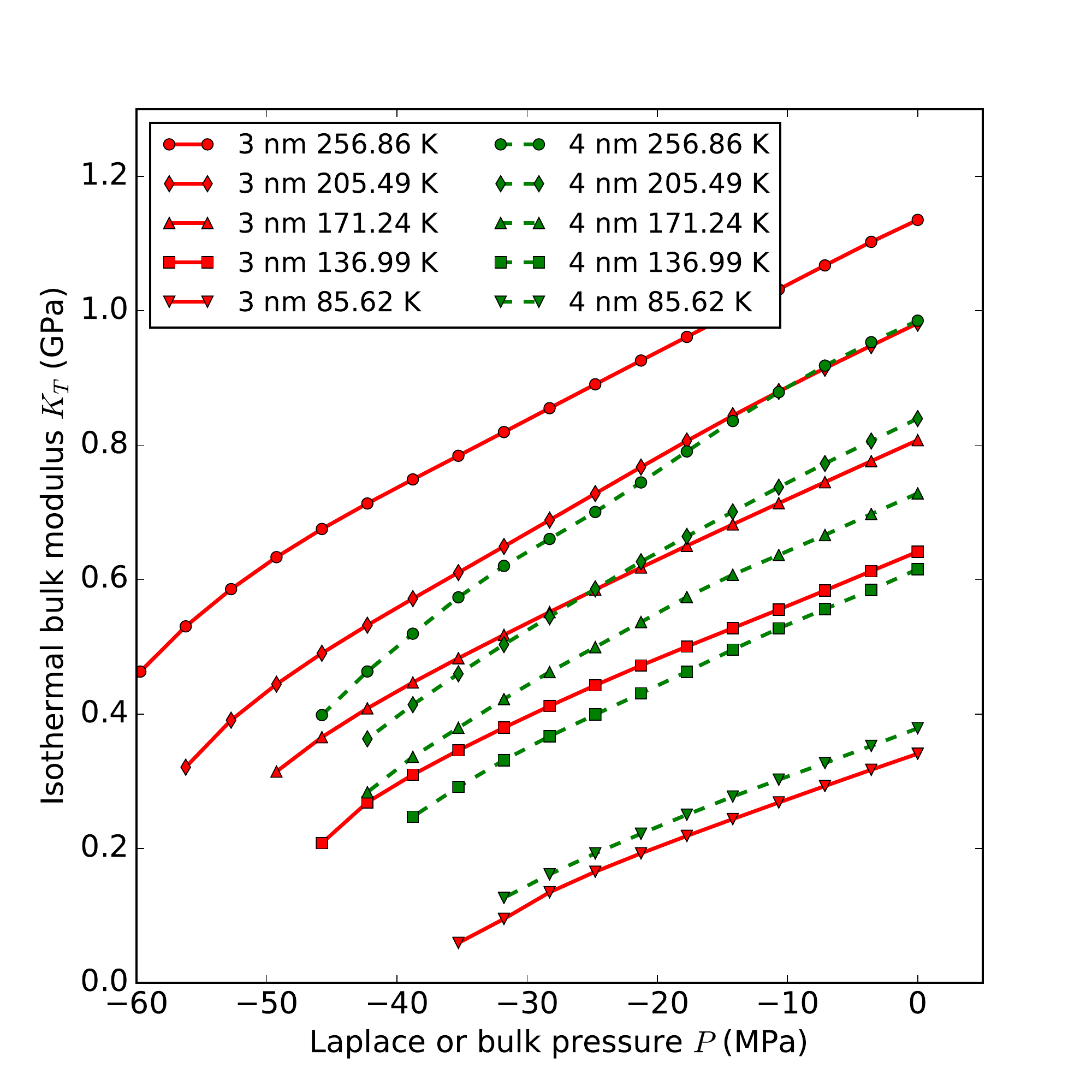} \\
\caption{Isothermal elastic modulus of liquid argon $K_T$ at 87.3~K as a function of Laplace pressure $P$, with varying solid-fluid interaction strength. Solid lines (red) are for external diameter 3~nm and dash lines (green) are for 4~nm. Solid-fluid interaction parameter $\epsilon_{\rm{sf}}/k_B$, is altered $-50\%$, $-20\%$, $0\%$, $+20\%$, and $+50\%$ relative to the original value of 171.24~K. The values of $\epsilon_{\rm{sf}}/k_B$ for each of the lines are given in the legend. 
}
\label{fig:Material}
\end{figure}

Similarly to our simulations results, Fig.~\ref{fig:Argon} also presents the $K_T$ moduli derived from ultrasonic experiments on argon confined in the Vycor glass nanopores from the recent works of Schappert and Pelster~\cite{Schappert2014, Schappert2014Langmuir} at 80 K and 86 K. We divided the adiabatic data from those papers by $\gamma = 1.969$ and $\gamma = 2.064$ for 80 K and 86 K, respectively~\cite{Tegeler1999}. Both series are shown as stars in Figure~\ref{fig:Argon}. 

Figure~\ref{fig:Argon} also includes the isothermal modulus $K_T$ for bulk argon as a function of bulk fluid pressure $p$. The horizontal dotted line gives the $K_T$ calculated at the saturation point $p = p_0 \simeq 0.1$~MPa at $T=87.3$~K. This value of $K_T$ agrees well with the experimental value $K_T = 0.47$~GPa~\cite{Tegeler1999}.  The positively sloped line with square markers gives the simulation results for liquid bulk argon for pressures above and below the saturation pressure at 87.3~K. Below the saturation conditions, the bulk liquid is necessarily metastable and, eventually, the fluid pressure becomes negative, indicating that the fluid is stretched. Finally, three solid lines in Figure~\ref{fig:Argon} show the $K_T$ for the bulk reference data at four temperatures 85~K, 87.3~K, 90~K and 95~K (top to bottom)~\cite{Tegeler1999,Linstrom_NIST_WebBook}. 

All the data sets shown in Figure~\ref{fig:Argon} (bulk and confined, theoretical and experimental) reveal one similar feature: a nearly linear dependence of $K_T$ on pressure, either the fluid pressure for a bulk liquid or the Laplace pressure for a confined fluid. We fit the data in Figure~\ref{fig:Argon} with the linear functions
\begin{equation}
\label{linear}
K(P) = K_i + \alpha P.
\end{equation}
We used all of the data points when fitting the experimental data for bulk and confined argon. When fitting the simulation data we use the data points above $-10$~MPa, which is the range of pressures comparable to the experimental data for confined argon. The linear fit is shown with the dashed lines of corresponding colors. The resulting fitting parameters are summarized in Table~\ref{table}. Except for the experimental ultrasonic data for confined argon at 86~K from~\cite{Schappert2014Langmuir}, which is very noisy, all the other series show very close slopes of $9.0 \pm 1.0$.

\begin{table}[ht]
 \begin{center}
    \begin{tabular}{| c | c | c | c | }
  \hline
 System & $T$ (K) & $\alpha$ & $K_i$ (GPa) \\ 
\hline 
Bulk (exp.) & 85.0 & 9.471 &  0.502 \\ 
Bulk (exp.) & 87.3 & 9.247 &  0.471 \\ 
Bulk (exp.) & 90.0  & 9.085 &  0.435 \\ 
Bulk (exp.) & 95.0 &  9.080 & 0.372 \\  
in Vycor (exp.) & 86.0 & 6.467 &  0.568 \\  
in Vycor (exp.) & 80.0 & 8.854 &  0.689 \\  
Bulk (GC-TMMC) & 77.7 &  9.304 &  0.605 \\
Bulk (GC-TMMC) & 87.3 &  9.293 & 0.475 \\
Bulk (GC-TMMC) & 95.7 & 9.127 & 0.373 \\
3 nm (GC-TMMC) & 77.7 & 8.961 &  0.979 \\
3 nm (GC-TMMC) & 87.3 & 8.941 &  0.823 \\
3 nm (GC-TMMC) & 95.7 & 8.282 &  0.718 \\
4 nm (GC-TMMC) & 77.7 & 8.481 &  0.867 \\
4 nm (GC-TMMC) & 87.3 & 8.541 &  0.729 \\
4 nm (GC-TMMC) & 95.7  & 9.662 &  0.626 \\
5 nm (GC-TMMC) & 87.3 & 8.798 & 0.672 \\ 
    \hline  
  \end{tabular}
 \end{center}
    \caption{Slope $\alpha$ and intercept $K_i$ (at $P = 0$) for the data shown in Figures~\ref{fig:Argon} and~\ref{fig:Temperature}.
   \label{table}    
    }
\end{table}

To further examine how this linear trend depends on temperature, Figure~\ref{fig:Temperature} shows $K_T$ for simulated argon in both bulk conditions and 3 and 4~nm spherical pores, at temperatures of 77.7~K, 87.3~K, and 95.7~K. Figure~\ref{fig:Temperature} again shows that $K_T$ is a nearly linear function of the appropriate pressure descriptor, particularly for pressures above $-10$~MPa. One feature of note is a pronounced ``hump'' centered about $-20$~MPa in the 4~nm, 77.7~K data set. This feature is simply statistical noise as uncertainty in $K_T$ is about 2\% and the linear trend line falls within 2\% error bars. The slopes for linear fits of the $K_T$ data in Figure~\ref{fig:Temperature} are also included in Table~\ref{table}. Remarkably, the slope of $K_T$ versus pressure for simulated argon falls between in the region $9.0 \pm 0.7$, irrespective of temperature or confinement.

Figure~\ref{fig:Density} shows the relationship between $K_T$ and the average fluid density for simulated argon as a bulk fluid and confined in the 3, 4, and 5~nm pores, and from experimental measurements of bulk argon, all at 87.3~K. All data in the figure are for liquid or liquid-like conditions (i.e., gas-like states for the confined fluid are not included). The common trend is that $K_T$ is a monotonically increasing function of density, and confinement increases $K_T$ relative to the bulk fluid. We will return to this plot for specific discussion in the following section.

Figure~\ref{fig:Material} shows the effect of the solid-fluid interaction on $K_T$ as a function of $P$ for spherical pores of size 3 and 4~nm, the intent of which is to show that the linear scaling trend is not specific to the argon-silica system originally studied, but extends to other materials (the interaction strength serving as a surrogate for material type). As in Figures~\ref{fig:Argon} and \ref{fig:Temperature}, $K_T$ is roughly linear with Laplace pressure near 0~MPa, though larger $\epsilon_{\rm{sf}}$ (i.e., strong solid-fluid interaction) shifts the entire trace of $K_T$ versus $P$ to higher values, and vice versa. The slopes of $K_T\left(P\right)$ for Figure~\ref{fig:Material} near $P=0$ are shown in Table~\ref{table:Material}. For all cases except the weakest interaction, that slope is again bounded by $9.0 \pm 1.0$, though monotonically increasing with $\epsilon_{\rm{sf}}$. For $\epsilon_{\rm{sf}}/k_B = $ 85.62~K (50\% reduction relative to the base argon-silica case), the slopes are close to 7.0. Additionally, there is an inversion in the modulus as a function of the solid-fluid interaction; for this case, the $K_T$ for the 4~nm pore is larger than that for 3~nm where the opposite is true for the other values of $\epsilon_{\rm{sf}}$. We note for this case that the solid-fluid interaction is weaker than the fluid-fluid interaction ($\epsilon_{\rm{ff}}/k_B$ = 119.6~K~\cite{Gor2015compr}), so the confinement is effectively solvophobic\cite{Gelb_Phase_1999}.

\begin{table}[ht]
 \begin{center}
    \begin{tabular}{| c | c | c | c | }
  \hline
 System & $\epsilon_{\rm{sf}}/k_B$ (K) & $\alpha$ \\ 
\hline 
3 nm   & 85.62 & 6.799 \\
3 nm   & 136.99 & 8.091 \\
3 nm   & 171.24 & 8.941 \\
3 nm   & 205.49 & 9.469 \\
3 nm   & 256.86 & 9.720 \\
4 nm   & 85.62 & 7.192 \\
4 nm   & 136.99 & 8.168 \\
4 nm   & 171.24 & 8.541 \\
4 nm   & 205.49 & 9.488 \\
4 nm   & 256.86 & 9.782 \\
    \hline  
  \end{tabular}
 \end{center}
    \caption{Slope $\alpha$ (at $P = 0$) calculated from GC-TMMC data as a function of solid-fluid interaction strength for 3~nm and 4~nm pores, based on the data shown in Figure~\ref{fig:Material}. 
   \label{table:Material}    
    }
\end{table}

\section{Discussion}
\label{sec:discussion}

The linear relation between $K_T$ of a bulk fluid or a bulk solid and the applied pressure (Eq.~\ref{linear}) has been known for decades. It is often referred to as the ``modified Tait equation''~\cite{Macdonald1966} or Tait-Murnaghan equation, originating from the seminal work of Murnaghan~\cite{Murnaghan1944}. Eq.~\ref{linear} is obviously a linearization of the $K(P)$ dependence. However, the range in which linearization works is very wide: for organic liquids it works well up to pressures ca.~100~MPa and for water up to ca.~1~GPa~\cite{Hayward1967}.

For bulk liquids the slope $\alpha$ in the $K(P)$ relation has been also discussed. Wilhelm pointed out that for a number of organic liquids this slope is practically independent of the temperature, and almost independent of the fluid, always about $\alpha \simeq 9-10$~\cite{Wilhelm1975}. Wilhelm also proposed an analytic expression for the slope $\alpha$ based on the Carnahan-Starling equation of state~\cite{Carnahan1969} with a cohesive term~\cite{Wilhelm1975}. His calculations were in good agreement with the experimental data for hydrocarbons. Using Wilhelm's theory we calculated the slope $\alpha$ for argon, using 3.48 \AA~ as a hard sphere diameter~\cite{Wilhelm1974} and the reference data for the triple point from~\cite{Tegeler1999}, yielding $\alpha = 8.2$, which is fairly close to the values of $\alpha$ calculated directly from reference data (Table~\ref{table}). We have to note that $\alpha$ is not a universal constant: its value for water (ca. $5.8$~\cite{Wilhelm1975}) and most solids~\cite{Anderson1966} is significantly lower.

Although one of us~\cite{Gor2014} had previously pointed out that $K_T$ has a linear dependence on pressure and the slopes agreed well with the experimental data from Ref. \onlinecite{Schappert2014}, we did not compare the confined fluid results to those of a bulk fluid. In the present work, we aggregate the results of simulations and experiments for both bulk and confined argon at multiple temperatures, and are thus able to conclude that, irrespective of whether the fluid is bulk or confined, its isothermal elastic modulus approximately satisfies a linear $K(P)$ relation (Tait-Murnaghan equation) with slope $\alpha$ that is only weakly dependent on either temperature or the nature of confinement. Therefore, our results show that nanoconfinement, while strongly affecting $K_T$ of the fluid (essentially, making the fluid stiffer), does not affect its pressure derivative $dK_T/dP$, the parameter $\alpha$ in the linear relation Eq.~\ref{linear}. We note that it is not the confinement \textit{per se} that increases $K_T$; the modulus is increased due to strong solid-fluid attraction forces. In the case of solid-fluid interactions that are weaker than fluid-fluid interaction (e.g.\ water in hydrophobic confinement), the opposite effect is observed: the value of $K_T$ in confinement is lower than the modulus of the bulk fluid~\cite{Strekalova2011, Evans2015, Evans2015PRL, Nygaard2016}.

As a general rule, higher density makes a fluid stiffer, i.e., increases the modulus $K_T$. That is indeed true for each of the data series in this work considered independently. However, the data in Figure \ref{fig:Density} clearly show that although the confined fluid has lower density than liquid argon (near saturation) at 87.3~K, it has a higher modulus.
Therefore, increased average density is not sufficient to explain the increase in $K_T$ for a confined fluid. Specifically, we note in Figure~\ref{fig:Density} that the 3~nm pore has generally lower average density than the other pores, yet has the highest modulus. It is likely that the reason for higher modulus of the confined fluid is due to contribution of the first 1-2 layers of fluid next to the pore wall. These layers are known to be solid-like, therefore the ``local'' modulus for these layers should be much higher than for molecules near the pore center. Experimental data for $K_T$ of solid argon at 77~K is ca. 1.4~GPa~\cite{Stewart1968, Anderson1975, Utyuzh1983}, i.e., larger than $K_T$ for liquid argon  by a factor of three. For smaller pores, the contribution of these solid-like layers to the overall modulus is higher and therefore the modulus of fluids in small pores significantly exceeds the bulk liquid value~\cite{Gor2015compr}. Unlike the modulus itself, the slope $\alpha$ of $K(P)$ dependence for solid argon is very close to that of the liquid,  $\alpha = 9.95$~\cite{Utyuzh1983}. This small difference explains our observation for confined fluids: while the modulus is noticeably higher for a given average density (due to the contribution of surface layers), the slope $\alpha$ is the same, because the solid-like layers still have the same $\alpha$.

The simulation results in Figure~\ref{fig:Material} test whether this common $\alpha$ extends to other materials for a given fluid. That figure clearly indicates that a linear trend near $P=0$ is preserved regardless of solid-fluid interaction strength and the $\alpha$ parameters in Table~\ref{table:Material} fall in the same bounds ($9.0 \pm 1.0$) provided that the confinement is solvophilic ($\epsilon_{\rm{sf}} > \epsilon_{\rm{ff}}$). This further suggests that the $\alpha$ parameter can be treated as a thermodynamic property of the fluid, not of the confining material, confinement dimensions, or temperature. Table~\ref{table:Material} also shows another new trend, that $\alpha$ appears to increase monotonically with solid-fluid strength, and prompts the question of whether further increasing $\epsilon_{\rm{sf}}$ would take $\alpha$ outside the range of $9.0 \pm 1.0$. We do not expect a further increased $\epsilon_{\rm{sf}}$ to alter $\alpha$ outside this range as packing constraints on the fluid near the material surface will prevent excessive densification of the confined fluid. Conversely, the weakening of $\epsilon_{\rm{sf}}$ to solvophobic conditions does lead to $\alpha$ outside the range of the bulk fluid. This is, however, expected due to the altered physics of the solvophobic confinement, in which the confined fluid is rarefied near the surface (compared to solvophilic confinement) and behaves more like a coalesced liquid than a typical capillary-confined fluid\cite{Striolo_Simulated_2004,Kimura_Cluster_2004,Liu_Does_2005,Strekalova2011,Strekalova2012,Evans2015}.

Our findings for the $\alpha$ of confined fluids further extend the broad commonality of this thermodynamic parameter, widely discussed in the past for the bulk fluids~\cite{Wilhelm1975}. In addition to the fundamental interest, this result imposes an important constraint on the analysis of ultrasonic data for fluid-saturated porous materials. Effective medium approaches, be it the one proposed by Schappert and Pelster~\cite{Schappert2013JoP}, or the one based on Biot-Gassmann equations~\cite{Gassmann1951, Biot1956i}, suggest that there is a direct proportionality between the change of the measured longitudinal modulus of the porous sample $\Delta M$ and modulus of the fluid $M_f$ (e.g., Eq.~\ref{EM}). The proportionality coefficient $C$ in this relation depends on the elastic properties of the dry porous sample, and the properties of the solid walls. While the former can be easily measured, the latter are hard to measure directly, and indirect methods for estimating them may be non-trivial and involve certain assumptions about the microstructure~\cite{Scherer1986, Gor2015modulus}. However, the general result of this work, i.e., the common slope $\alpha$ of the $K(P)$ line, imposes a constraint on Eq.~\ref{EM}, and unambiguously determines the value of the constant $C$ in this relation. Therefore, one can readily compare the slope of the experimentally measured dependence of $\Delta M = \Delta M (p/p_0) = \Delta M (P_L)$ to the slope $\alpha$ of the reference $K_T$ data for the same fluid and to calculate the translation coefficient $C$ from this comparison. Analysis of the value of $C$ based on an effective medium approach will then provide the information of the elastic constant of the pore walls (solid constituent). 

\section{Conclusion}

We have studied the elastic modulus-pressure dependence $K(P)$ for confined and bulk liquid argon. For bulk fluids and solids this dependence is well described by a linear function $K(P) = K(0) + \alpha P$, known as Tait-Murnaghan equation. We have used transition-matrix Monte Carlo simulations to calculate the isothermal elastic modulus $K_T$ of argon in bulk and confined in silica mesopores. Our calculations have shown that although $K_T$ is strongly affected by confinement, its dependence on pressure can still be described by the linear Tait-Murnaghan equation. The calculated slope $\alpha$ for confined argon is the same as $\alpha$ calculated for bulk one provided that the confinement is not solvophobic; moreover, it agrees well with the reference data for bulk argon and experimental data for confined argon derived from ultrasonic experiments. Also it does not appreciably changes with temperature. In addition to shedding more light on thermodynamics of confined fluids, our results advance the analysis of ultrasonic experiments on fluid-saturated porous samples. Knowledge of the slope $\alpha$ of the $K(P)$ dependence can be applied as a constraint to the analysis of experimental ultrasonic data, making it possible to estimate the elastic properties of an unknown porous medium. 

\section*{Acknowledgment}

This research was performed while one of the authors (G.G.) held a National Research Council Research Associateship Award at Naval Research Laboratory. The work of G.G. and N.B. was funded by the Office of Naval Research through the Naval Research Laboratory's basic research core program. G.G. thanks Rolf Pelster and Klaus Schappert for the fruitful discussions of ultrasonic data, George Scherer for discussion of mechanical properties of Vycor glass, and Patrick Huber for the argon heat capacity data.

\end{document}